# Uncovering the system vulnerability and criticality of human brain under dynamical neuropathological events in Alzheimer's disease

Jingwen Zhang[1], Qing Liu[2], Haorui Zhang[2], Michelle Dai[3], Qianqian Song[4], Defu Yang[5], Guorong Wu[5,6], and Minghan Chen[1]

**Author affiliations:**

[1] Department of Computer Science, Wake Forest University, Winston-Salem, NC 27109, USA

[2] Department of Mathematics, University of North Georgia, Oakwood, GA 30566, USA

[3] Department of Biostatistics, Harvard T.H. Chan School of Public Health, Boston, MA 02115, USA

[4] Department of Cancer Biology, Wake Forest School of Medicine, Winston-Salem, NC 27101, USA

[5] Department of Psychiatry, University of North Carolina at Chapel Hill, Chapel Hill, NC 27599, USA

[6] Department of Computer Science, University of North Carolina at Chapel Hill, Chapel Hill, NC 27599, USA

**Corresponding author and lead contact:** Minghan Chen

Full address: 1834 Wake Forest Road, Winston-Salem, NC 27109

E-mail: chenm@wfu.edu

## ABSTRACT

*Background:* Despite the striking efforts in investigating neurobiological factors behind the acquisition of amyloid-β (A), protein tau (T), and neurodegeneration ([N]) biomarkers, the mechanistic pathways of how AT[N] biomarkers spreading throughout the brain remain elusive.

*Objectives:* To disentangle the massive heterogeneities in AD progressions and identify vulnerable/critical brain regions to AD pathology.

*Methods:* In this work, we characterized the interaction of AT[N] biomarkers and their propagation across brain networks using a novel bistable reaction-diffusion model, which allows us to establish a new systems biology underpinning of Alzheimer's disease (AD) progression. We applied our model to large-scale longitudinal neuroimages from the ADNI database and studied the systematic vulnerability and criticality of brains.

*Results:* Our model yields long term prediction that is statistically significant linear correlated with temporal imaging data, produces clinically consistent risk prediction, and captures the Braak-like spreading pattern of AT[N] biomarkers in AD development.

*Conclusion:* Our major findings include (i) tau is a stronger indicator of regional risk compared to amyloid, (ii) temporal lobe exhibits higher vulnerability to AD-related pathologies, (iii) proposed critical brain regions outperform hub nodes in transmitting disease factors across the brain, and (iv) comparing the spread of neuropathological burdens caused by amyloid-β and tau diffusions, disruption of metabolic balance is the most determinant factor contributing to the initiation and progression of Alzheimer's disease.

**Keywords:** Alzheimer's disease, AT[N] biomarkers, brain network, reaction-diffusion model, vulnerable and critical regions

# INTRODUCTION

Alzheimer's disease (AD), a progressive neurological disorder, is documented by post-mortem examination or by biomarkers *in vivo,* according to the National Institute of Aging and Alzheimer's Association (NIA-AA) [1]. The biomarker framework proposed by NIA-AA for clinical diagnosis includes: (i) extracellular plaques consisting of amyloid-β (Aβ, referred to as A biomarker) [2–4]; (ii) intracellular neurofibrillary tangles (NFT) that are the intraneuronal aggregation of hyperphosphorylated tau protein (referred to as T biomarker) [5,6]; and (iii) neurodegeneration (referred to as [N] biomarker) that is characterized by neuronal loss and exhibits metabolic, structural, and functional deficiency [7].

According to the amyloid hypothesis, amyloid-β triggers a series of downstream pathological responses, including inflammatory responses, hyperphosphorylation of tau protein, and AD progression [8,9]. Tau protein helps to stabilize microtubules, but when hyperphosphorylated, it disassembles from axons, aggregates into neurofibrillary tangles, blocks neuron-to-neuron transportation, and consequently leads to the loss of synaptic functions and neurons [5,6]. The toxicities of pathological amyloid and tau lead to malfunctioning neurons, which further induces neurodegeneration, the molecular manifestation of brain atrophy or malfunctioning neurons in both normal aging brains and neurodegenerative diseases including Alzheimer's disease [5,7,10].

Recently, both clinical and computational findings support the trending mechanistic hypothesis that the spread of AT[N] biomarkers exhibits a specific temporal and diffusive pattern [11–13]. Increasing attention has been paid to the spatial and temporal correlation between tau depositions and neuron losses. Some researchers favor the tau hypothesis [5,6,14] over the amyloid hypothesis [15,16] as amyloid is poorly matched to the spatial distribution and temporal evolution of neuron loss. Yet, the abnormal transformation of tau is shown to be mediated by amyloid-β [17,18]. Moreover, several studies emphasize the necessity of both amyloid-β and tau in AD development by showing that tau does not lead to AD in the absence of amyloid accumulation [9,10]. In this context, it is crucial to study the interaction between amyloid and tau and their dynamic impact on the pathophysiological mechanism of AD.

The recent advance in neuroimaging techniques offers a window to measure the pathological burden and structural atrophy *in vivo* along with the progression of AD. Currently, most neuroimaging studies utilize association-based approaches to understand the neurobiology risk factors behind AD progression. Particularly, consistent efforts have been made in deep learning fields to better predict AD progression. In 2016, Hosseini-Asl et al. used a deep 3D convolutional neural network (CNN) to learn features from magnetic resonance imaging (MRI) data and to differentiate between AD patients and healthy controls [19]. In another study by Zhao et al., a graph convolutional network (GCN) was used to analyze brain connectivity data and identify changes in the brain's structural network associated with AD [20]. Recently, Luo et al. (2021) utilized a variational autoencoder (VAE) to learn features from multimodal neuroimaging data and predict the progression of AD. These studies achieved high accuracy in differentiating AD patients from healthy controls and identified several brain regions that were strongly associated with disease progression. However, even with intricate and convoluted models, more training data and rounds of validations/testing are needed before we could possibly integrate them into clinical practice due to the black-box nature of machine learning approaches. The lack of a system-level understanding could potentially lead to findings that are distinct from essential physiopathological mechanisms. In this regard, the pioneering network-diffusion model [11,21] was used to predict longitudinal atrophy patterns from MRI images. The recent epidemic spread model [18,22] investigated the spread of amyloid and tau on structural and functional networks. However, those models only describe the diffusion process of disease factors while ignoring the fundamental interactive pathways between AT[N] biomarkers. Although tremendous efforts have been made to model complex biological systems, most AD-related systems biology approaches are limited to studying a single pathological pathway or a small part of the brain, lacking the whole-brain insight gained from the large-scale longitudinal neuroimaging data [23,24].

This work aims to understand the pathophysiological mechanism of AD by probing into the spatiotemporal dynamics of AT[N] biomarkers on the whole-brain scale. We conceptualize that AD-related biomarkers not only interactively contribute to the neurodegenerative process at each brain region but also influence the connected regions in a prion-like manner. To this end, we deploy a network-guided bistable model to characterize the AT[N] cascade interactions and diffusion patterns (Fig. 1B). The longitudinal neuroimaging data is used as the benchmark to evaluate the predicted evolutionary trajectory, including (i) regional AT[N] biomarker concentration levels extracted from PET scans and (ii) structural brain networks constructed from T1-weighted MRI and diffusion-weighted imaging (DWI) scans (see the *Input* in Fig. 1A). We also include the cognitive reserve proxy [25] to model the resistance to neuropathology burden. The model converges to two stable steady states (shown in the cyan dash box of Fig. 1A): the low-energy cognitive normal state (low-risk state, LRS) and the high-energy AD state (high-risk state, HRS), which lays the cornerstone for prediction risk, a key indicator of AD likelihood.

Model outcomes present an overall strong linear correlation between prognostic and diagnostic results (see the *Output* in Fig. 1A) and capture the Braak-like spread pattern of AT[N] biomarkers in AD development. Furthermore, based on system behaviors manifested in the reaction-diffusion model, we are able to identify brain regions that suffer great vulnerability to the abnormal AT[N] burdens, as well as to nominate a collection of nodes that reveal critical importance to pathological progression across the brain under complex neuropathological events. By providing mathematical insights into the mechanism of Alzheimer's disease and its spreading pattern, our model could assist the development of new therapies to slow or halt disease progression.

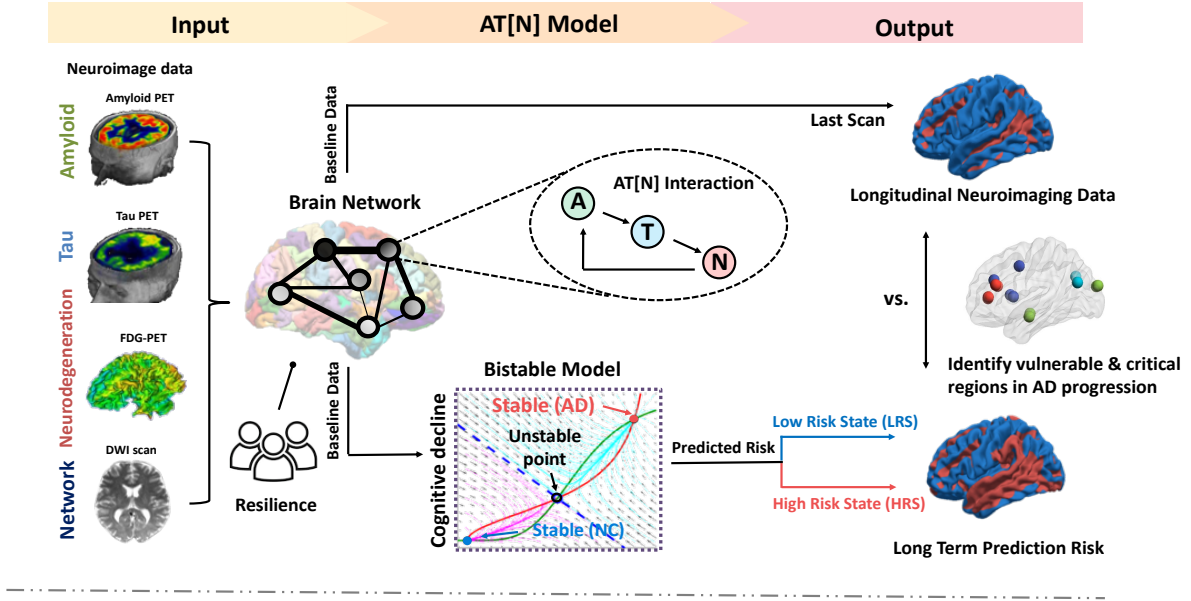

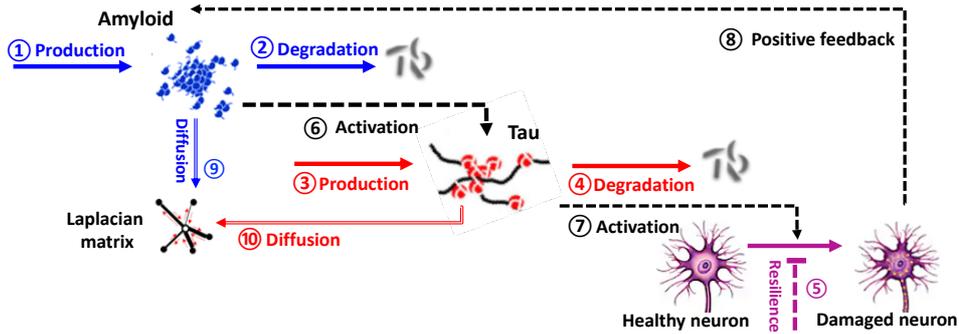

**Figure 1. Overview.** (**A**) Methodological process. *Input:* Amyloid-, Tau-, FDG-PET, DWI scans, and cognitive reserve proxy are processed and used as indicators of regional AT[N] biomarker level, network connectivity, and neuropathological resistance, respectively. *AT[N] Model:* our reaction-diffusion model is built on the AT[N] cascade mechanism and characterizes bistable states (low-, high-risk states) of the system. *Output:* the last scans of regional neurodegeneration levels of each subject are classified into low/high states using the optimal cutoff value found from receiver operating characteristic analysis by maximizing the sum of sensitivity and specificity, which are then used to train the model parameters. The model outputs long-term prediction risk trained, represented as a surface rendering of regional low/high risks and identifies brain regions that are (1) vulnerable to pathologic burden and (2) critical in transmitting biomarkers across the brain. (**B**) Bistable reaction-diffusion model. The backbone of our proposed model is built on the interactive pathways and neuronal prion-like propagation hypothesis of AT[N] biomarkers: amyloid-β activates the hyperphosphorylation of tau protein, and the abnormal tau triggers neurodegeneration which then leads to AD; both amyloid and tau spread across the brain network in a prion-like manner. See *Materials and Methods* section for detailed descriptions of mechanistic pathways.

# MATERIALS AND METHODS

## Participants

All data used in this study were leveraged from Alzheimer's Disease Neuroimaging Initiative (ADNI) database. In total, 1616 subjects from ADNI passed our quality control after image processing and parcellation. Among them, 320 participants were selected to train the model based on three criteria: (i) have Amyloid-PET, Tau-PET, FDG-PET, T1-weighted MRI, and DWI scans; (ii) have at least one follow-up PET scan of A, T, or [N]; (iii) have a clinical diagnostic label (cognitive normal or AD) for each PET scan. Note that FDG data (a measure of tissue glucose metabolism) is used as a reversed indicator of neurodegeneration. Region-to-region connectivity strength, measured by the count of white matter fibers, is also utilized to characterize brain network diffusion.

## Data processing

PET data. Each PET scan (Amyloid-PET, Tau-PET, FDG-PET) was aligned with their own subject's T1-weighted MR image. For each brain region, we calculate its standardized uptake value ratio (SUVR) to represent the pathological burden at each cortical region. We divide the tracer uptake in the region of interest by the uptake in the whole gray matter volume, which has relatively low tracer accumulation and is considered to be unaffected by the disease being studied. This normalization helps to account for variations in tracer administration and other factors and allows for comparison of tracer uptake across different individuals and studies. In order to make the SURV measurement more robust, we use the bootstrapping procedure to adaptively sample the point and calculate the region-wise average of SUVR [26].

Brain network construction. Using the software of FreeSurfer v5.6 [27], all MRI data were processed through four steps: (i) skull stripping; (ii) tissue segmentation into white matter, gray matter, and cerebrospinal fluid; (iii) cortical surface reconstruction based on tissue segmentation map; and (iv) cortical surface parcellation into 148 regions of Destrieux atlas [28] using deformable image registration. We then aligned the DWI images to the corresponding T1-weighted MR image for each subject. Following the parcellation of cortical surface, we applied surface seed-based probabilistic fiber tractography in FreeSurfer v6.4.0.5 with "probtrackx" and "bedpost". Thus, each element in the structural network is essentially the fiber count, where the total number of fiber counts varies from subject to subject. Considering such individual differences, we normalize the connectivity matrix such that the connectivity degree is invariant to individual brains. Since it is more interpretable to understand the connectivity degree associated with each brain region as a probability, we apply row-wise normalization instead of whole brain normalization. Given the normalized connectivity matrix $\mathbf{W}$, we further make it symmetric by $\mathbf{W} = \frac{1}{2}(\mathbf{W} + \mathbf{W}^T)$. Thus, we calculate the Laplacian matrix by $\mathbf{L} = \mathbf{D} - \mathbf{W}$, where $D$ is a diagonal matrix of node-wise connectivity degree.

Resilience proxy. In this work, we calculated population-wise estimated resilience proxy, the ability of an individual to resist the cognitive decline associated with AD. This estimation is based on a mathematical model using subjects' demographic data, socioeconomic factors, cerebrospinal fluid (CSF) tau/Aβ ratio, and AD-related polygenetic risk scores, which we found this interaction term manifests the role in counteracting the progression of AD in our statistical model [25].

*Reaction-diffusion model*

Our proposed network-guided biochemical model consists of a classic bistable model and network diffusion. This relatively simple model enabled us to investigate the spatiotemporal dynamics of AT[N] biomarkers in AD by capturing the essence of the underlying mechanism of complex biological phenomena. In Fig. 1B, there are five major entities in our model: (i) A biomarker (written as $x_A$), representing the Aβ protein, can be measured from Amyloid-PET [29,30]. (ii) T biomarker (written as $x_T$), representing the tau protein, can be measured from Tau-PET [31]. (iii) [N] biomarker (written as $x_N$), measured from MRI or FDG PET, is an indicator of neuronal injury which is simplified as the damage caused by A and T biomarkers [32–34]. $x_A$, $x_T$, and $x_N$ are three column vectors assembling the observed degree of AT[N] biomarkers at each region. $x_A$, $x_T$, and $x_N$ are three column vectors assembling the observed degree of AT[N] biomarkers at each region. (iv) A $148 \times 148$ structural brain network matrix to represent the normalized region-to-region connectivity strength (written as Laplacian matrix $\mathbf{L}$). In the diagonal of matrix, we subtract the total connectivity degree of each region to reflect the inward and outward spreading of pathological burden throughout the brain network. (v) A $148 \times 1$ vector $r$ representing the population-wise averaged, regional specific cognitive reserves proxy which mediates and even delays the neurodegeneration process.

The pathological framework is an integration of AT[N] reactions and network diffusion including five main pieces:

1. Constant production of A and T (①, ③);
2. Density-based degradation/clearance of A and T (②, ④);
3. Regional network resilience to counteract the neurodegeneration(⑤);
4. Non-linear cascade activations: amyloid accelerates the accumulation of pathologic tau (A→T), which then activates neurodegeneration process (T→N) and in turn amyloid deposition through positive feedback pathway (N → A) (⑥, ⑦, ⑧);
5. Prion-like diffusion of A and T on the structural brain network (⑨, ⑩).

At each brain region, the production (①, ③) and clearance (②, ④) of amyloid and tau proteins are included in the model following zero-order and first-order mass-action kinetics, respectively. We also include the cognitive reserve proxy [25] to model the individual's network resilience (in terms of the moderated ratio of neuron loss), denoted as the inhibition pathway (⑤). The interaction of AT[N] biomarkers follows the dominant amyloid cascade hypothesis [35,36] and is denoted as activation pathways (⑥, ⑦). The phenomena of damaged neurons stimulating amyloid production via reactive astrocytes [37] are represented by the positive feedback pathway (⑧). The classic Hill function [38,39] is applied to approximate the multi-molecular interacting process in the activation and feedback pathways as nonlinear reactions. The part of AT[N] reactions constitutes a classic bistable model. Finally, the diffusion of amyloid and tau proteins along the white matter fiber pathways in the structural brain (⑨, ⑩) is modeled using the graph equivalent Laplacian matrix **L** as the diffusion operator on the brain network, where **L** characterizes the dynamic balance of influx and outflux of neuropathological burdens at each node. The diagram in Fig. 1B can be converted into three sets of PDEs (Eqs. 1), which model the spatiotemporal dynamics for A, T, and [N] biomarkers, respectively.

$$\begin{cases} \frac{\partial x_A}{\partial t} = \underbrace{k_{pA}}_{①} - \underbrace{k_{dA} x_A}_{②} + \underbrace{k_{NA} \frac{(x_N)^\alpha}{K_{MN}^\alpha + (x_N)^\alpha}}_{⑧} + \underbrace{d_A \mathbf{L} x_A}_{⑨} \\ \frac{\partial x_T}{\partial t} = \underbrace{k_{pT}}_{③} - \underbrace{k_{dT} x_T}_{④} + \underbrace{k_{AT} \frac{(x_A)^\beta}{K_{MA}^\beta + (x_A)^\beta}}_{⑥} + \underbrace{d_T \mathbf{L} x_T}_{⑩} \\ \frac{\partial x_N}{\partial t} = \underbrace{-k_\gamma r \cdot x_N}_{⑤} + \underbrace{k_{TN} \frac{(x_T)^\gamma}{K_{MT}^\gamma + (x_T)^\gamma}}_{⑦} \end{cases} \quad (1)$$

We use hyperparameter $\Theta$ to denote model parameters which are necessary to characterize the rate constants of production ($k_{pA}$, $k_{pT}$), clearance ($k_{dA}$, $k_{dT}$), activation and positive feedback ($k_{NA}$, $k_{AT}$, $k_{TN}$), inhibition ($k_r$), diffusion ($d_A, d_T$), dissociation ($k_{MA}, k_{MT}, k_{MN}$), and the coefficients in Hill function ($\alpha, \beta, \gamma$). Our PDE-based model can be used to predict the evolution of AT[N] biomarkers given the baseline biomarkers, and understand the complex physiopathological mechanism of AD by analyzing the system behaviors as described next.

*System stability analysis*

By solving the characteristic equations of our PDEs, we can find the equilibria of our model, which captures a nonlinear dynamical system on a continuous-time domain. Lyapunov's stable theory [40] is applied to further analyze the local stability of the detected equilibria. The equilibrium found is stable if and only if the real part of the solution to the characteristic equations or eigenvalue of the Jacobian matrix are all negative. Our bistable system generates two stable equilibria, mirroring the diagnosis of Alzheimer's disease: the small equilibrium represents the low-risk state with low accumulation of AT[N] biomarkers, and the high-risk state with high biomarker levels. The characterization of stability analysis allows us to predict the spatiotemporal evolution of AT[N] biomarkers and investigate vulnerable structures in the brain where a subtle disturbance would significantly influence the dynamics of other regions.

*Parameter optimization*

The receiver operating characteristic (ROC) analysis was used to facilitate the classification of disease status using biomarker SUVR values as the correlation between cognition and biomarker is not collinear [41,42]. Subjects were first divided into two groups based on their clinical labels: the low-energy cognitive normal (CN) state (including CN, SMC, and EMCI diagnostic labels) and the high-energy AD state (including LMCI and AD diagnostic labels). An optimal cutoff value was found for each brain region by maximizing the sum of sensitivity and specificity. The optimal cutoff values we found for each biomarker then served as a threshold to classify individuals' regional SUVR value into either positive (+) or negative (-) states. We use AT[N]'s respective positivity as an optimization objective to achieve data-driven classification, aiding in long term prediction. The median accuracies are 74.06%, 73.75%, and 62.97% for A, T, and [N] biomarkers, respectively. The median sensitivity and specificity are 64.81% and 78.54% (A), 51.85% and 85.85% (T), 58.33% and 65.57% (N) with a median area under the curve value of 0.7599 (A), 0.7139 (T), and 0.6387 (N).

In the simulation, all neuroimaging data were scaled into [0,1] to represent the relative regional concentration of AT[N] biomarkers. The model takes the regional AT[N] biomarkers as input and outputs the estimated risk for every region of interest [28]. The bistable attribute of our model enables it to generate simulation results: each brain region evolves into either a low-risk state or a high-risk state, where LRS indicates regions predicted to remain healthy and HRS indicates regions predicted to accumulate abnormal AT[N] burden and develop into AD lesions. The average of LRS and HRS are considered as the "prediction risk" discussed in our model. Three different algorithms (Genetic Algorithm [43], Bayesian Algorithm [44], and Direct Search [45]) were tested to optimize the hyperparameter $\boldsymbol{\Theta}$. While all three optimization algorithms return comparable results, GA has a faster convergence rate and returns a slightly smaller difference between the neuroimage classification and simulation results.

*Statistical analysis*

All statistical analyses were performed using R. Pearson correlation coefficient was calculated to measure the degree of linear correlation between two sets of data. Two-tailed student's t-tests were used to test the significance of correlation coefficients. Two-tailed Student's t-tests were used for single comparisons between two sets of data. Statistical significance was concluded with *p*-value ≤ 0.05. 95% confidence level is used for confidence intervals. Results in tables are presented as mean ± standard error of the mean.

# RESULTS

*Subject information*

Among all processed subjects' data, we selected 320 subjects spanning the AD spectrum, where selected subjects have longitudinal neuroimaging scans of Amyloid-, Tau-, FDG-PET, and DWI. Based on their diagnostic labels, 241 subjects are categorized into CN group and 79 subjects labeled as LMCI or AD were grouped into AD group. Since the correlation between cognition and biomarker is not collinear [41,42], we adjust the diagnostic label by taking biomarker's positivity into consideration in result analysis. We used 1.5 interquartile rule (IQR) to identify outliers in Amyloid-, Tau-, and FDG-SUVR, and adjusted the label as AD/CN accordingly: high outliers in CN group are adjusted to AD state, and low outliers with AD diagnosis are adjusted to CN state. The labels now take both clinical symptoms and biomarkers positivity into account. See *Materials and Methods* section for sample selection criteria and imaging processing specifications.

*System stability*

Since a high reaction rate usually indicates a rapid disease progression, it is fundamental to investigate when the disease will evolve into a stable stage (remains in CN or AD), which can be characterized by the system time it takes to reach equilibria (steady state) from the initial condition. Like the previous setting, we track the stability time under different rate constants of production, clearance, activation, diffusion, and resilience for amyloid and tau. Compared to amyloid, tau presents more variations in stability time when the production rate is tuned up by 50% (Fig. 2A), and the clearance rate is down by 20% (Fig. 2B). We also observe a positive association between tau activation rate and stability time (Fig. 2E), illustrating the vital role of tau hyperphosphorylation in AD progression. The resilience rate shares the same trend with the clearance rate but poses significantly more influence on stability time (Fig. 2F). Increasing resilience significantly decreases the time needed to reach a stable state. This result, together with the discussion on predicted risk, implies the importance of non-biological factors, such as education and social interaction, play an important role in slowing AD progression [25,46,47].

Diffusions of amyloid and tau do not change the characteristic time of system stability. Diffusion rates of both amyloid and tau are the least influential parameters on stability time, which implies alterations in structural brain networks are not the key driving force of AD progression despite the diffusive nature of AT biomarkers (Fig. 2C, G). We further examine the influence of structural networks on disease progression. Although some outliers affect the fitting performance, we observe positive associations between connectivity strength and stability time (the system time it takes to reach a steady state from the initial condition) across both CN and AD groups in Fig. 2D, which means nodes with lower connectivity strength tend to reach stable states faster. The dynamics of network diffusion alone can also be characterized by the second eigenvalue of the graph Laplacian matrix, denoted as $\lambda_2$, which dominates the convergence speed to steady state or system equilibrium. Fig. 2H displays the bootstrap distribution of "relaxation time" ($1/\lambda_2$), an essential measurement of the time scale for a system to return to its steady state after a perturbation. We could see a clear stratification between the relaxation time for CN and AD groups. A larger relaxation time (as seen with the AD group) indicates a longer diffusion process and therefore may lead to longer-lasting damage due to increased duration of amyloid or tau presence.

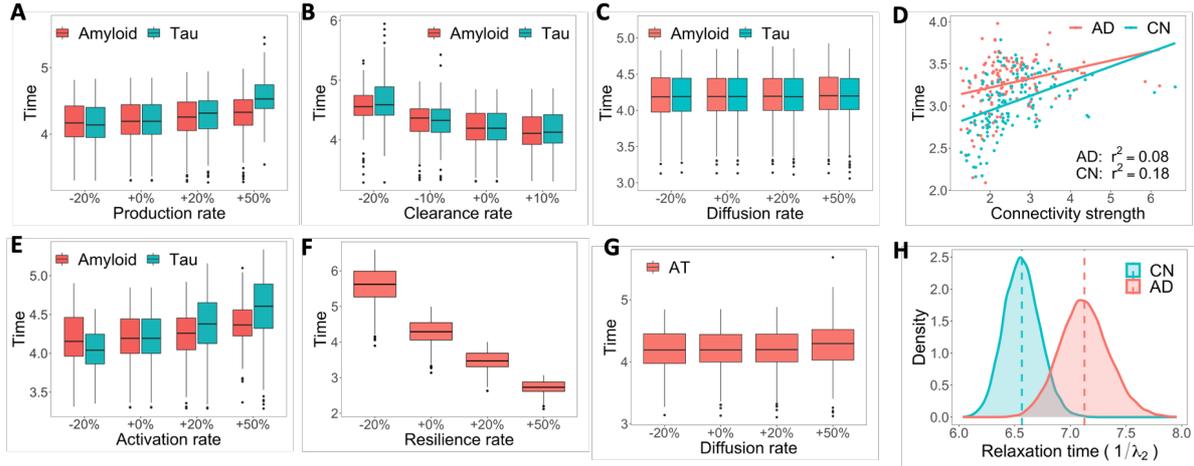

**Figure 2. Influence of amyloid and tau production, clearance, activation, and diffusion rates on stability time. (A)-(C), (E)-(G)** Boxplots of stability time with varied rate constants of production, clearance, diffusion, activation, and resilience. The *x*-axis represents the relative increase (+) or decrease (-) of rate constants. +0% is the base reaction rate used in the model. The y-axis represents the time that the system takes to reach a stable state. (**D**) Linear regression fitting of connectivity strength over stability time (the time to reach system equilibrium) for each diagnostic group. (**H**) Bootstrap distribution of relaxation time (inverse of the second eigenvalue of Laplacian matrix, or convergence speed to system equilibrium) for CN and AD groups.

*Braak-like spread pattern*

Inspired by a wealth of preclinical *in vivo* and *in vitro* research in the past decade that suggests amyloid and tau proliferation in the brain follows a prion-like pattern transmitting between neurons [21,48,49], we use this general spread pattern to examine the validity of our model. Amyloid and tau progression were divided into three stages (Stages Early, Middle, Late) based on prevailing amyloid staging [50,51] and tau staging [52,53]. In the early stage of tau as shown in Fig. 3H, the mild affection of neurofibrillary tangles and neuropil threads (NT) are confined to the entorhinal cortex. Adjacent limbic areas including amygdala and para-hippocampal cortex and temporal cortices are impacted by NFT and NT at middle stage. At the late stage, all part of the hippocampal formation is attacked and NFT with long extension reaches the outer part of the iso-cortex, and finally reaches primary cortex areas (precentral gyrus). The progression of amyloid falls into a relatively reversed pattern compared with tau (see Fig. 3D). In the early stage of amyloid progression, amyloid hits regions including precuneus, medial orbitofrontal cortices, and posterior cingulate, and then spread out into almost all iso-cortical associated regions in the middle stage. The late stage is characterized by loads of amyloid in lingual, sensorimotor, and nearby regions.

As we can see from Fig. 3A-C, amyloid first deposits over fronto-medial and temporo-basal areas and disperses into the adjacent iso-cortex. Amyloid then severely affects the remaining associative neocortex. It arrives at the striatum at the final stage, leaving the entire brain affected. For amyloid, our model successfully classifies 68 regions as middle stage and all regions in late Braak stage, while misclassified 10 regions in middle stages to late stage and misclassified 14 regions as CN state, which achieves 83.8% accuracy. We correctly predicted some well-documented neocortical regions such as precentral gyrus, precuneus, etc. Fig. 3E-G has revealed a similar spread pattern as Braak staging of tau. Tau first spreads from entorhinal and trans-entorhinal regions to para-hippocampal, fusiform, amygdala, and related regions. From there, tau accumulates across the entire iso-cortex and brain. For tau, our model successfully classifies all regions in middle Braak stage, 78 regions as late stage, and leaves 28 regions in low energy state, which achieves 81.1% accuracy. Our model successfully captures the stage of important regions such as entorhinal cortex, angular gyrus, and inferior temporal lobe.

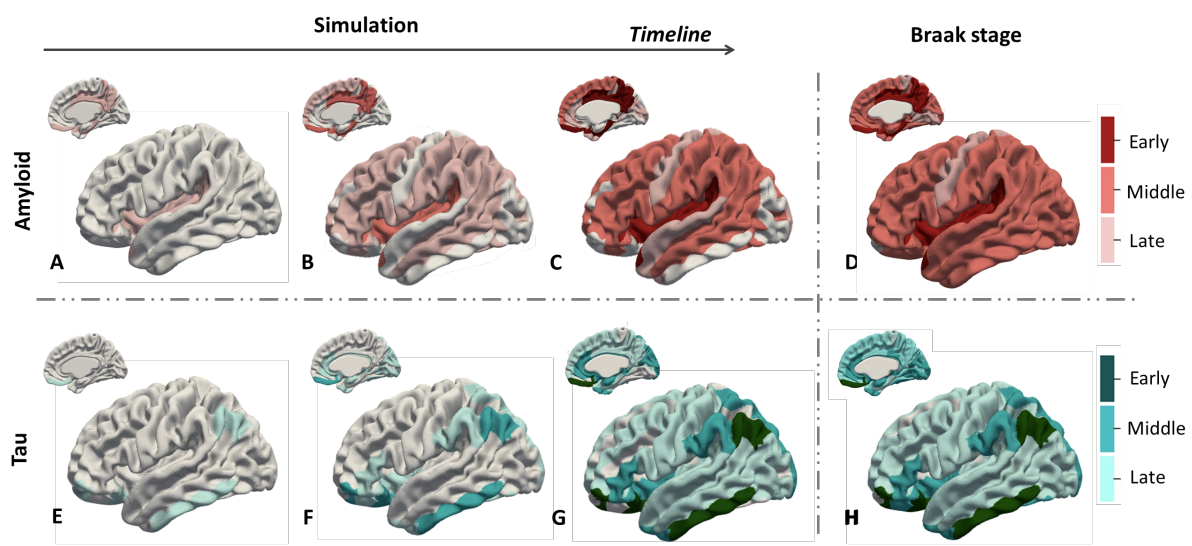

**Figure 3. Spread pattern of amyloid and tau in our model and Braak stages**. (**A**), (**E**) Initial onset of abnormal amyloid (red) and tau (green). (**B**), (**F**) Progressive pattern of AT at the middle stage of simulation. (**C**), (**G**) Final pattern of AT accumulation across the brain at the end of simulation, when the system arrives at the stable state. (**D**), (**H**) Prion-like transmitting pattern between neurons of AT proposed by Braak. The dark to light spectrum of red/green represents early, middle, and late stages of amyloid/tau Braak stage.

*Model performance*

Overall model outcomes will be evaluated with AT[N] biomarkers and clinical cognitive assessment from different aspects and significant associations are found between them, demonstrating our models' potential to uncover the heterogeneous progressive pattern of AD. We will examine individual predicted risk (proportion of high-risk state for each subject across the entire brain (148 nodes)) and regional predicted risk (proportion of high-risk state for each brain region across all subjects).

Regional tau and neurodegeneration are stronger indicators of regional AD risks. In Fig. 4A-C, as we averaged our patients' data, this last scan data presented the severity of each biomarker on all levels of patients. We observed severe amyloid burden in the inferior temporal gyrus, superior temporal sulcus, as well as cingulate gyrus and precuneus (Fig. 4A). Minimal amyloid deposition in inferior frontal gyrus. For tau, high concentrations of tau from patients were detected in the temporal gyrus, and low concentrations of tau were observed in paracentral lobule (Fig. 4B). For neurodegeneration biomarkers, severe levels of neurodegeneration encompass the temporal gyrus and nearby regions, and fewer neurodegenerations were observed in posterior-dorsal part of the cingulate gyrus (dPCC) (Fig. 4C). Fig. 4D was based on the predicted risks generated by our model, which is an indicator of the severity of AD biomarkers' progression. Extreme levels of predicted risk appear to depose at inferior temporal gyrus and middle temporal gyrus, and low predicted risk regions are posterior-dorsal part of the cingulate gyrus as well as inferior frontal gyrus. These results align well with the current understanding that tau and neurodegeneration are more closely related to AD progression in space and time [54,55].

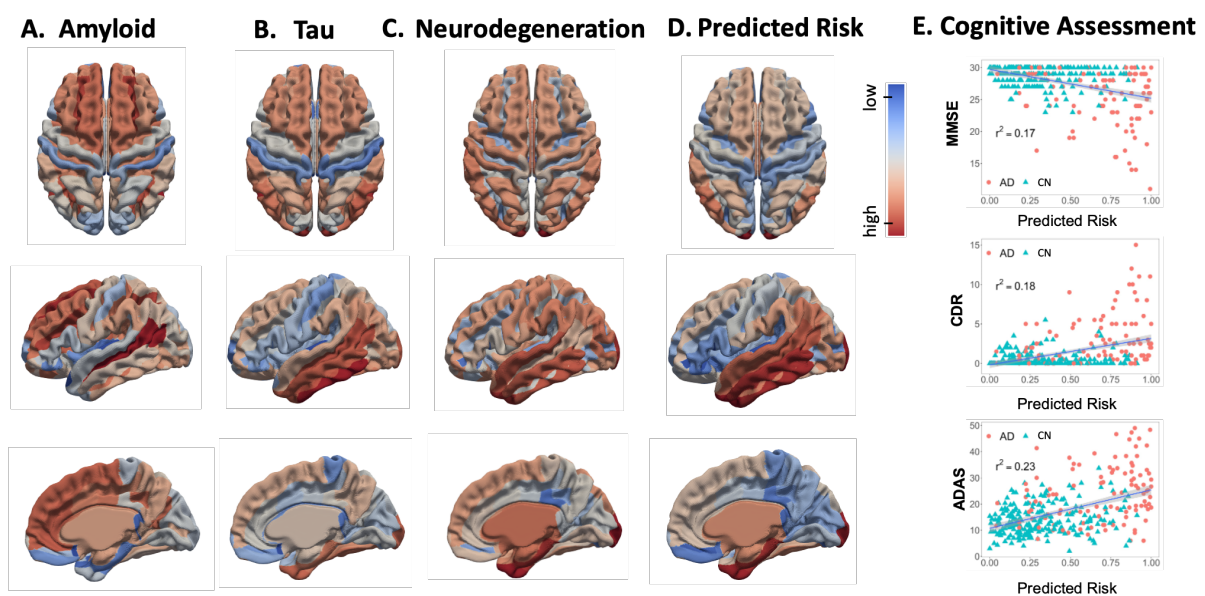

**Figure 4. Model performance in prediction risk.** (**A-C**) Surface rendering of beta-amyloid, pathologic tau, and neurodegeneration averaged over all 320 subjects in axial and sagittal views. The color spectrum ranges from red (high concentration) to blue (low concentration). (**D**) Surface rendering of prediction risk averaged over all 320 subjects in axial and sagittal views. The color spectrum ranges from red (high risk) to blue (low risk). (**E**) Individual predicted risk vs. cognitive assessments (MMSE, CDR, and ADAS), with colors and shapes indicating adjusted labels.

Our predicted risk accords with averaged AT[N] patterns with an apparent increase of prediction risk in the middle and inferior temporal gyrus from the sagittal view of the brain, demonstrating the vulnerability of the temporal lobe and the indicative sign in early AD diagnosis [11,56]. We further measure the validity of our model using clinical cognitive assessments (Fig. 4E). Our prediction risks are found to be linearly associated with Mini-Mental State Exam (MMSE) scores ($r^2 = 0.17, 95\% \, CI = [0.10, 0.24]$), Clinical Dementia Rating (CDR) scores ($r^2 = 0.18, 95\% \, CI = [0.11, 0.25]$), and Alzheimer's Disease Assessment Scale (ADAS) ($r^2 = 0.23, 95\% \, CI = [0.15, 0.31]$), as shown in Fig. 4E.

Amyloid and tau progression patterns affect model prediction. Our primary model performance presents a strong linear relationship between AT[N] neuroimaging examinations and cognitive tests. Here we investigate our model results by examining the predictive error. By fitting the least square linear regression model of regional neurodegeneration [N] over average regional risk, regions of interest are then classified as (i) fitted prediction if the observed [N] lies within the 99.9% confidence interval (CI) of predictions, (ii) overestimations if the observation lies below the CI of predictions, (iii) underestimations if the observation lies above the CI (Fig. 5A). A similar pattern is detected across AT[N] biomarkers when we compare the surface rendering of neurodegeneration (Fig. 5E) with the average level of amyloid and tau over 320 subjects (Fig. 5B, F). Our model tends to overestimate the risk of AD for regions with high A and T profiles (red regions in Fig. 5B, F) and underestimate the risk of AD at regions with low A and T profiles (cyan color regions). Fig. 5C and G show the significant difference between AT levels in overestimated group and underestimated group, which confirms that overestimated regions have higher amyloid and tau burden than underestimated areas. The model residual shows a negative correlation with observed amyloid ($r^2 = 0.11$) and tau deposition ($r^2 = 0.32$), implying that the aberrant regional amyloid and tau deposition affects the model accuracy in AD risk evaluation (Fig. 5D, H).

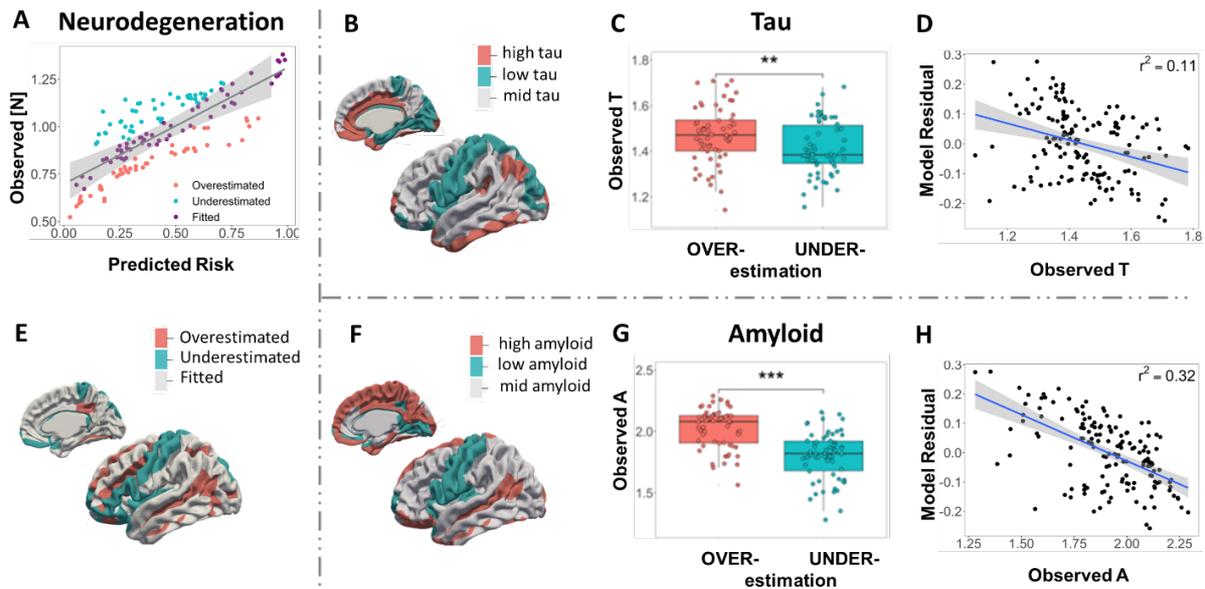

**Figure 5. Amyloid and tau explain regional model under-, over-estimation. (A)** Predicted risk vs. observed [N] biomarker (last scans averaged over subjects). Brain nodes are classified as overestimated or underestimated according to the sign of model residual. **(E)** Surface rendering of under/overestimated neurodegeneration. **(B), (F)** Amyloid and tau deposition pattern averaged over subjects' last scans, with red representing a high level and blue representing a low level using the optimal cutoff value. **(C), (G)** Boxplots of regional amyloid and tau level in under/overestimated groups. **(D), (H)** Correlations between regional model residual and biomarkers tau, amyloid. Negative model residuals indicate model overestimations, while positive model residuals indicate model underestimations.

### *Vulnerable regions in AT[N] pathological progression*

Studies have shown that AT[N] biomarkers tend to affect different areas of the brain. We summarized brain regions that are frequently reported as being affected in literature for A, T, and [N] biomarkers, respectively [57–64]. In general, higher levels of amyloid deposition were found in the anterior cingulate, frontal cortex, lateral temporal cortex, parietal cortex, precuneus, and anterior ventral striatum in participants with mild cognitive impairment compared with normal control [57,61,64]. Regional Tau-PET levels revealed that tau affected trans-entorhinal cortex, entorhinal cortex, and medial temporal limbic system heavily except for the hippocampus [59,61,63]. Regions with the drastic decline in cortical thickness ([N]) exhibit in the left anterior cingulate, dorsolateral prefrontal cortex, orbitofrontal cortex, visual association cortex, and medial temporal lobe, outlining para-hippocampal gyrus [58,60,62]. See *Supplementary* for more detailed descriptions and the brain mapping of summarized regions.

Temporal and occipital lobes suffer vulnerability to pathological AD progression. To unveil brain regions vulnerable to abnormal AT[N] burdens, we perform an extensive simulation of random abnormal onset on the brain. By randomly placing abnormal malignant AT[N] onsets across brain regions, we record the nodes that reach HRS (high-risk state) after system stabilization. A group of 14 nodes is identified as most susceptible to developing into HRS regardless of the initial abnormal onset and are therefore referred to as vulnerable regions. As shown in Fig. 6B, 64.3% of vulnerable nodes locate in temporal and occipital lobes and 35.8% of them are among the worst 10% of neurodegenerative regions. Since HRS is assumed to be associated with high AT[N] profiles, we check the bar plot of regional AT[N] together with the predicted risk, sorted by the ascending risk (Fig. 6A). Concert with previous understanding, many nodes with higher risks have higher accumulations of amyloid, tau, and neurodegeneration. However, the AT[N] SUVR values of some vulnerable regions (such as nodes #21, #23, #41) are at a medium level, indicating that vulnerable regions do not necessarily refer to regions with the most neuropathological burdens. We further compare the longitudinal changes (Δ) in AT[N] level and predicted risk of proposed vulnerable nodes and non-vulnerable nodes (Fig. 6C). The increments of amyloid and tau in identified vulnerable regions ties to non-vulnerable regions, but vulnerable regions bear significant more neural loss compared to non-vulnerable regions $p$-value= 0.005. The predicted risk also reflects the same trend, where vulnerable regions are associated with significantly higher predicted risk compared to other regions ($p$-value= 0.0001). Further comparison of $\Delta$AT[N] between vulnerable nodes shows that the overall increase of neuropathological burdens at vulnerable regions are higher than regions summarized from literature [57–64], especially for tau (See *Supplementary* Fig. S1 for details). Together with the analysis of predicted risk, our results confirm the susceptibility of vulnerable nodes and significant increases of neurodegeneration level in AD development.

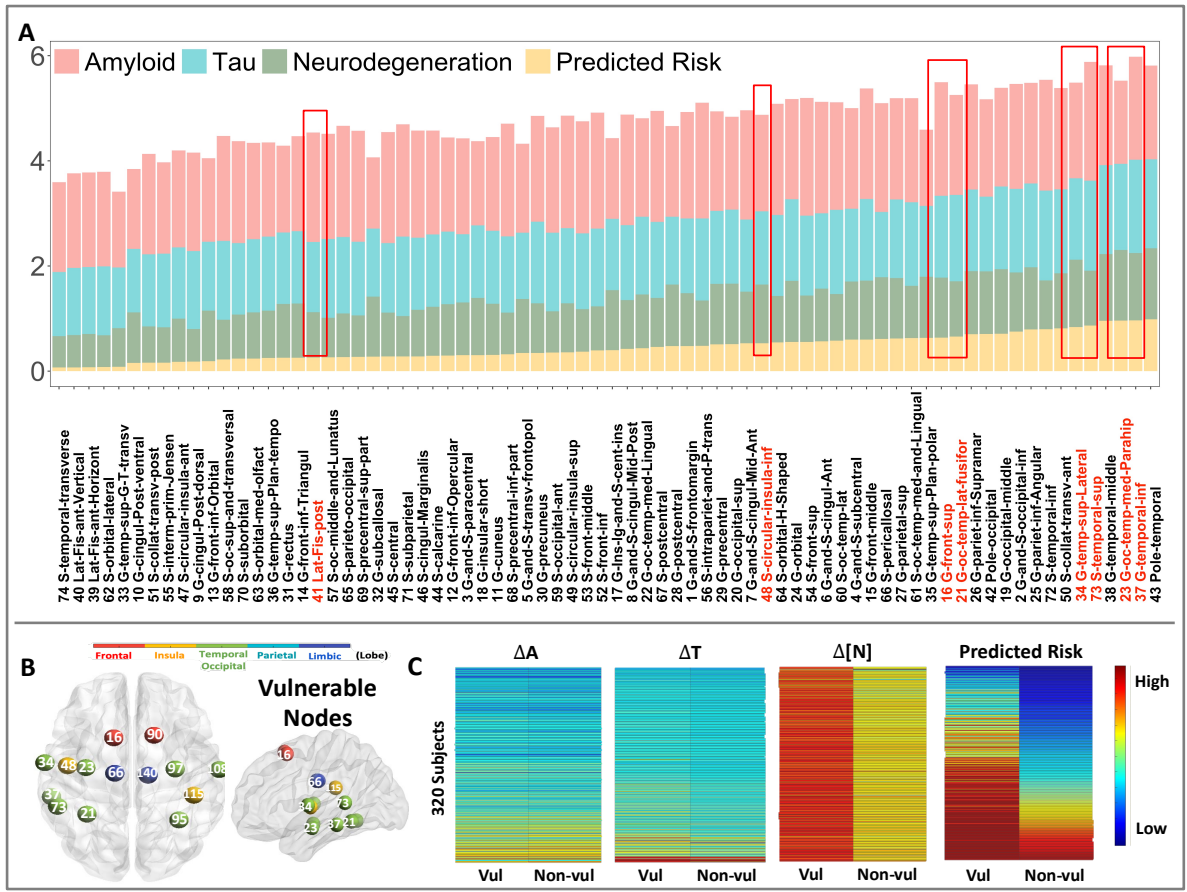

**Figure 6. Vulnerable regions to AT[N] burden.** (**A**) A stacked bar plot of AT[N] SUVR values (y-axis) and predicted risk for 74 parcellated brain regions in the left hemisphere (x-axis), sorted in ascending order of predicted risk. Vulnerable nodes are highlighted in red rectangles. (**B**) Brain mapping of vulnerable regions identified by our model. Ball colors indicate different brain lobes. Node IDs are written in white, referring to region name in (**A**). (**C**) Heatmaps of average longitudinal changes of AT[N] between last scan and first scan for vulnerable vs. non-vulnerable regions from 320 subjects. All heatmaps are sorted by the predicted risks of non-vulnerable nodes in descending order.

## Critical regions of AT[N] transmission network

Besides vulnerable regions that are susceptible to abnormal AT[N] burdens, we are also interested in regions that act as hubs in integration and potentially augment metabolic cascades relevant to brain disease. Considering the diffusive nature of AT biomarkers, these regions are expected to be transmissible to disease factors, where subtle increases can quickly spread out and significantly affect neighboring and further areas. In this section, we proposed a set of critical nodes based on our model result in comparison with the prevailing hub nodes from the intersections of current research studies [65,66], in hope of shedding light on precision medicine and early prediction in AD field.

Regional onset of critical nodes outperforms hub nodes. We summarized a set of hub nodes obtained from in silico research on both functional network and spatial network of AD patients. They have been identified to serve as gateways for information processing and communication. Hub nodes cover transentorhinal entorhinal cortex, posterior cingulate, and fusiform gyrus, which are two of the ROIs with dense anatomical and functional connections to many other brain regions. For the selection of critical brain regions, we increase the AT[N] levels for random groups of nodes, count the number of brain regions that end up at HRS after system stabilization, and choose the set of nodes that convert most brain regions into HRS. Our Critical Nodes are proposed from model performance in understanding the regions that could drive most changes in AT[N] level, which involve precuneus, superior frontal gyrus, precentral gyrus, supramarginal gyrus, lateral superior temporal gyrus, olfactory sulcus, and inferior temporal gyrus. In both sets, we found inferior temporal gyrus to be crucial in propagating AD burdens. However, our model presents us with possible signal centers such as precentral gyrus and precuneus, which are highly correlated with early AT depositions and substantial pathological burdens in the late stage of AD progression.

Fig. 7 presents the comparison between our proposed critical nodes (Fig. 7C) with hub nodes (Fig. 7B) from literature [65–67] and a set of randomly selected nodes (Fig. 7A), where hub nodes and critical nodes coincide in the region of bilateral orbital gyri. We (1) tune the three sets of nodes (blue nodes in Fig. 7A-C) to abnormal levels, (2) set the rest of brain regions to normal AT[N] levels, (3) run the simulation until the system reaches its stable state, and (4) evaluate the transmissibility of disease factors for those nodes by counting the number of regions that progress to abnormal (AD) stage. Critical nodes impact and drive as many as 49 nodes to HRS (Fig. 7F), whereas hub nodes and random nodes affect 32 nodes (Fig. 7E) and 24 nodes (Fig. 7D), respectively. The predicted risks of critical nodes are larger than 0.7 in the left and right inferior temporal gyrus, left and right superior temporal gyrus, and left supramarginal gyrus. Our finding is consistent with the previous study that temporal gyrus, which plays a major role in object recognition, is one of the cognitive functions that are impaired early on in AD [68,69].

Further, we investigate the graph-theoretic metrics: degree, PageRank, and closeness centrality for each region of interest (ROI) across all subjects. In Fig. 7G-I, we compare the overall degree (a measure of connectivity strength), PageRank (an identification of influential nodes whose influence extends beyond their direct connections), and closeness centrality (a measure of how long it will take to spread information sequentially from the current node to all other nodes). We observe a clear stratification between critical nodes, hub nodes, and random nodes for degree (Fig. 7G) and PageRank (Fig. 7H): critical nodes have notably higher degrees and PageRank compared to hub nodes, which are both higher random nodes. The majority of critical nodes have closeness higher than hub nodes and random nodes. But there is no distinct pattern between the comparison of closeness from hub nodes and random nodes. This indicates that the structural network is an important factor in determining AD transmission and progression.

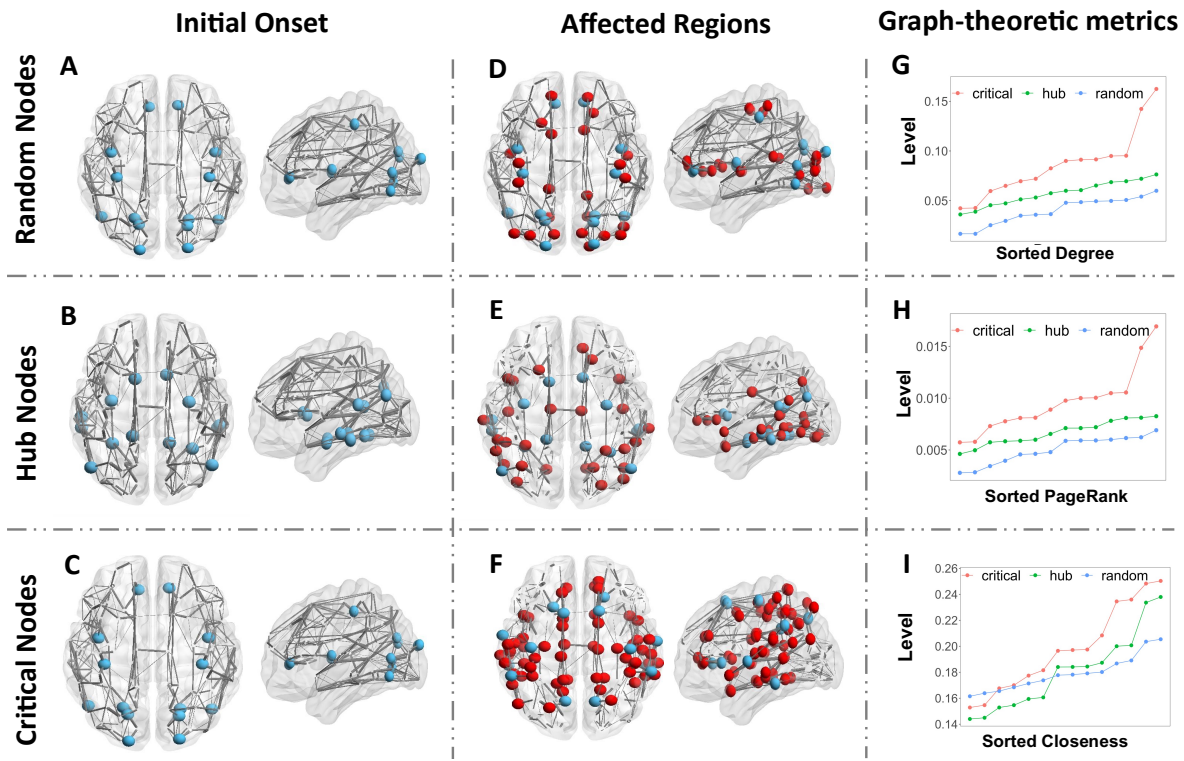

**Figure 7. Regional onset of critical brain regions compared with hub nodes and random nodes.** (**A**)-(**C**) Initial regional onset of randomly selected nodes, hub nodes, and critical nodes. (**D**)-(**F**) Predicted brain regions under high risk using random nodes onset, hub nodes onset, and critical nodes onset. Blue balls indicate the location of random nodes, hub nodes, and critical nodes; red balls are brain regions affected and developed to high-risk state; grey lines represent an averaged brain network with its thickness denoting the relative region-to-region connectivity strength. (**G**)-(**I**) Graph-theoretic metrics of degree, PageRank, and closeness centrality (sorted in ascending order) at random nodes, hub nodes, and critical nodes, respectively.

*Pathway and network influence*

To better understand the influential roles of various pathways in our network-guided model, we examine the system behaviors under different amyloid and tau production, clearance, activation, and diffusion rates (See *Materials and methods* for detailed descriptions of mechanistic pathways).

Overproduction and clearance deficiency of amyloid and tau significantly increase AD risk. Minor abnormalities in amyloid and tau usually can be self-corrected by neurons through different cleaning mechanisms such as proteolytic degradation and out-of-brain transportation via blood-brain barrier [2,42]. However, AD patients are unable to restrain these disease factors within the normal range due to their disrupted production-clearance balance [22,70]. To test the clinical impact of this metabolic balance, we experimentally vary the rate constants of reactive pathways in our system and trace the changes in predictive risk for CN and AD groups. Despite a certain degree of overlapping, the predicted risk for the AD group is noticeably higher than the CN group, see Fig. 8A-H. We also observe a statistically significant difference in predicted risks among varied production rates and clearance rates of AT (Fig. 8A, B, E, F). The changes in amyloid and tau production rates explain about 4.5% ($r^2$) and 14.3% of the increase in predicted risk, respectively; the changes in amyloid and tau clearance rates explain 27.0% and 16.3% of the decrease in predicted risk. These suggest that elevated production rates and suppressed clearance rates will significantly increase the risk of AD.

Network diffusion is not a determinant factor in AD progression. We then check the effect of activation and diffusion for amyloid and tau. A similar stratified risk level is observed between different clinical labels: AD groups have the highest risk, followed by MCI and NC. For the activation pathway, a general rising pattern of risk is witnessed in both amyloid and tau as we increase the rate constant (see Fig. 8C, G). When increasing the activation rate by 50%, the risks from the three groups overlap, indicating that hyper-active amyloid and tau may lead to high risk regardless of subjects' diagnostic stage. The changes in amyloid and tau activation rate explain 12.3% and 5.0% of the increase in predicted risk, respectively. Even though network connectivity is recognized as a critical factor in AD, it is worth noting that there is no significant increase (0.006% and 0.0006%) in risk prediction (Fig. 8D, H). Scaling up the diffusion rate alone may accelerate the speed of disease progression but does not change the final disease stage, and thus is not influential to predicted risk here. Finally, we check the effect of the resilience rate (Fig. 8I). Increasing the resilience rate significantly decreases predicted risks for all clinical labels, and the change in resilience rate explains 10.3% of the increase in predicted risk. Our finding, concerting with other previous statistical studies [25,46,47], shows the effects of education and socioeconomic factors in delaying AD progression.

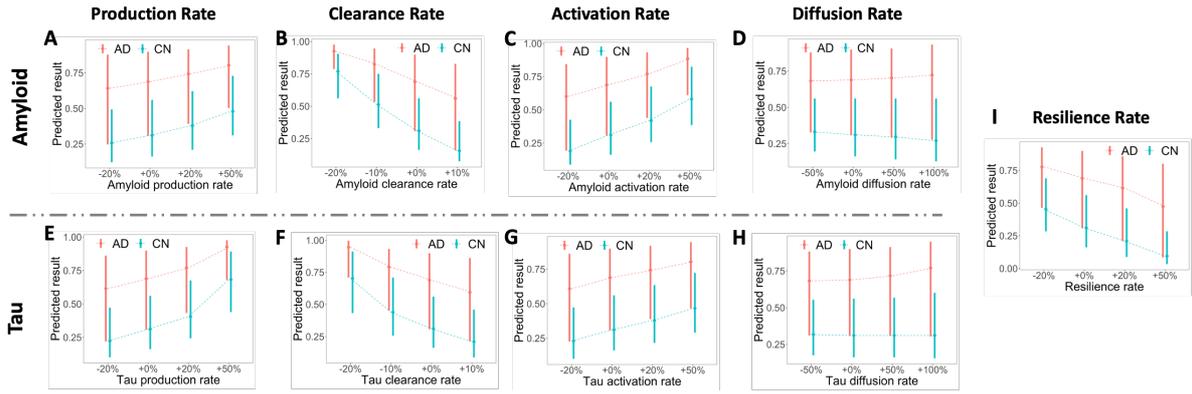

**Figure 8. Influence of amyloid and tau production, clearance, activation, diffusion, and resilience rates.** (**A**)-(**I**) Predicted risk with varied reaction rate constants. For each plot, bars indicate the 25 and 75 percentiles of predicted risk, and the dashed line connects the mean of predicted risk under different reaction rates. The *x*-axis represents the relative increase (+) or decrease (-) of rate constants. +0% is the base reaction rate used in the model. The colors denote the diagnostic labels of each group (CN, AD).

## DISCUSSION

Alzheimer's disease is irreversible and slowly progressive dementia with limited treatments. Due to the multiplicity of clinical symptoms, standard neuropsychological assessments inadequately reflect the underlying pathophysiological mechanisms, rendering a significant gap between neurobiological examinations of AD pathology and clinical diagnoses. The in-depth understanding of how AT[N] biomarkers spread throughout the brain is a crucial step because disentangling the regions that are vulnerable to disease factors and the regions that are highly "contagious" after suffering from abnormal accumulation of those neuropathological burdens is the gateway for precision medicine. Based on our previous model [71], we processed considerable neuroimaging data from two-fold individual subjects, added in longitudinal Tau-PET data, and replaced original MRI data with fluorodeoxyglucose PET for more accurate measurement of neurodegeneration. The baseline and final scans of Amyloid-, Tau-, and FDG-PET are used as initial and final indicators of AT[N] biomarker levels accordingly.

While AT[N] biomarkers are considered pathological hallmarks of Alzheimer's disease, none of them is unique to Alzheimer's disease. What makes the simulation more complex is that high-level amyloid, tau, and neurodegeneration accumulations are not limited to AD subjects but also exist in some normal aging populations. Thus, measurement of AT[N] level alone is insufficient for clinical diagnosis and a universal cutoff value for biomarker abnormality might be inaccurate for classification. A well-designed model is in demand to differentiate normal accumulation and pathological progression. In this study, we employ the proposed network-guided reaction-diffusion model to analyze the impact of biomarker interactions and network diffusion on AD progression. By using Braak staging as a benchmark, we evaluate the validity of our model based on its ability to replicate this widely accepted progression stages of Alzheimer's disease pathology based on the regional distribution of neurofibrillary tangles and neuritic plaques in the brain (Fig. 3). Since the simulation results are governed by the reaction-diffusion model, we further evaluate the correlation between the global attribute of system behaviors (the percentile of predicted high-risk regions) and cognitive measurements (CDR, ADAS, and MMSE), which all exhibits significant positive relationships ($p < 10^{-16}$), as shown in Fig. 4E. Such a noticeable association between system behaviors and clinical outcomes shows the potential of our model in disentangling the heterogeneity of neurodegeneration trajectories.

While our model results are accurate on a general view, over- or under-estimations do occur with high and low regional AT burdens. Certain nodes with higher regional amyloid and tau deposition tend to have low neurodegeneration which leads to an over-estimation by our model. The spatial concurrence of AT[N] biomarkers could explain this prediction error as abnormal amyloid and tau deposition is associated with and can accelerate neurodegeneration [9,54,55]. This urges the need to include more longitudinal AT[N] biomarkers in model construction as they provide more comprehensive information on the underlying microscale disease progression. Notice that the simplification of our model might also contribute to this estimation error. Our current model does not account for the "incubation time" from the build-up of amyloid to the onset of tau pathology, either to the onset of neurodegeneration or clinical dementia. To investigate the transition time from the build-up of amyloid to the onset of tau pathology and neurodegeneration, we first categorize the longitudinal biomarkers' burden for each subject by using receiver operating characteristic based on the optimal cutoffs that classify subjects' longitudinal scans as normal (−) or abnormal (+). Provided with the categorized profiles, we then apply logistic regression to model the longitudinal trajectories of AT[N] biomarker transitions. By doing this, we found that the average time lapse from ($A^-→A^+$) to ($[N]^-→[N]^+$) transitions could be as long as 18 months [54]. The inclusion of temporal latency in future modeling could potentially improve the overall accuracy of AT[N] biomarker evolution.

Extensive research has proven the detrimental effects of mitochondria functionality, oxidative stress, long-term potentiation, synaptic plasticity, and memory caused by the irregularity of amyloid-β [1,2]. Yet these potential disease factors can be maintained at a normal level through different clearance mechanisms, including proteolytic degradation, out-of-brain transportation via blood-brain barrier, immune responses, and protein-mediation [2,22,72]. If the metabolic balance is interrupted, either through overproduction or clearance deficiency, a chain of downstream consequences will be triggered, and eventually leads to AD. This conclusion is backed by our model results with tuned parameters of amyloid and tau. Increasing the production rate or decreasing the clearance rate can cause a notable rise in predicted risk. When reducing the clearance rates of amyloid or tau node by 20%, all regions surge in the high-risk range regardless of clinical diagnostic label, indicating that the interrupted metabolic balance may be a key to the initiation and progression of Alzheimer's disease [70,73,74]. Our model also indicates that the increased level of resilience shows effectiveness in delaying or preventing AD progression. In the current literature on cognitive reserve, there is converging evidence of the existence of resilience against the development of neuropathologies, which is highly dependent on the lifestyle factors such as education and occupation [25].

Recent research shows the critical role of brain network during the evolution of Alzheimer's disease. Several laboratory studies showed that hyperphosphorylated tau would spread along with structural brain networks, leading to neuron loss [11,48]. Our study, however, reveals that the final predicted risk will not be affected if we increase the diffusion rate alone. While it could be partially explained by the measurement error during neuroimaging acquisition and data processing, the static brain network could play a significant role here. In our model, we used the average network connectivity of CN and AD groups to simulate individuals' AT[N] dynamics. Since the neurodegenerative process could alter the network topology [27], network alteration might likely manifest a dynamic propagation of pathological burdens. We can incorporate subject-specific longitudinal brain networks in the study to further examine the influence of network alternation in disease progression.

With an extensive search via randomly seeding amyloid and tau disease factors, we found the temporal lobe, especially the middle and inferior temporal gyrus, suffers great vulnerability to abnormal AT disposition. Unlike traditional studies, our work shows that the brain regions that are affected most by abnormal AT[N] burdens are not constrained to regions with high amyloid or tau depositions. Some regions, such as the dorsolateral prefrontal cortex and - visual association cortex, bearing average AT burdens, are at high risk to neuropathological alternation caused by AD, encouraging more attention to those regions for clinical diagnosis and treatment. Besides vulnerability, our model also reveals that bilateral inferior temporal gyrus, bilateral superior temporal gyrus, precuneus, and superior frontal gyrus are exceptionally transmissible to AT[N] burdens during Alzheimer's progression. The criticality of those brain regions provides insights into a new interpretation of neuroimaging data in diagnosis and advocates early treatments on those particular sites to delay or prevent potential future whole-brain AD development.

Our analyses on empirical data reveal a concurrent progression of T[N] biomarkers and a strong indicative power of T[N] profiles in AD prediction. Based on the proposed systems biology model, we proved the importance of maintaining the metabolic balance of amyloid and tau in AD prevention and targeting temporal lobe in clinical intervention. More importantly, our findings of vulnerable regions that are severely affected by, and critical regions that are highly contagious to AT[N] profiles would shed light on early AD prediction and precision medicine. Future work considering delayed pathways and network alteration may further improve our model framework and accuracy.

## ACKNOWLEDGMENT

The authors have no acknowledgments to report.

## FUNDING

The work is supported by NIH R03AG073927 and Wake Forest University Pilot Grant Funding.

## DECLARATIONS OF COMPETING INTEREST

The authors declare that there is no conflict of interest.

## DATA AND CODE AVAILABILITY

The data that support the findings of this study are openly available in ADNI database at http://adni.loni.usc.edu/. The code and any additional information required to reanalyze the data reported in this paper are available from the lead contact upon request.

## REFERENCES

[1] Jack CR, Bennett DA, Blennow K, Carrillo MC, Dunn B, Haeberlein SB, Holtzman DM, Jagust W, Jessen F, Karlawish J, Liu E, Molinuevo JL, Montine T, Phelps C,

Rankin KP, Rowe CC, Scheltens P, Siemers E, Snyder HM, Sperling R, Elliott C, Masliah E, Ryan L, Silverberg N (2018) NIA-AA Research Framework: Toward a biological definition of Alzheimer's disease. *Alzheimers Dement* **14**, 535–562.

[2] Chen G-F, Xu T-H, Yan Y, Zhou Y-R, Jiang Y, Melcher K, Xu HE (2017) Amyloid beta: structure, biology and structure-based therapeutic development. *Acta Pharmacol Sin* **38**, 1205–1235.

[3] Hardy JA, Higgins GA (1992) Alzheimer's disease: the amyloid cascade hypothesis. *Science* **256**, 184–186.

[4] LaFerla FM, Green KN, Oddo S (2007) Intracellular amyloid-β in Alzheimer's disease. *Nat Rev Neurosci* **8**, 499–509.

[5] Ballatore C, Lee VM-Y, Trojanowski JQ (2007) Tau-mediated neurodegeneration in Alzheimer's disease and related disorders. *Nat Rev Neurosci* **8**, 663–672.

[6] Goedert M (1993) Tau protein and the neurofibrillary pathology of Alzheimer's disease. *Trends Neurosci* **16**, 460–465.

[7] Crews L, Masliah E (2010) Molecular mechanisms of neurodegeneration in Alzheimer's disease. *Hum Mol Genet* **19**, R12–R20.

[8] Kg M, W S, V O, L M, T K, Jc M, Ke Y, Rj B (2010) Decreased clearance of CNS beta-amyloid in Alzheimer's disease. *Science* **330**, 1774–1774.

[9] Musiek ES, Holtzman DM (2015) Three dimensions of the amyloid hypothesis: time, space and "wingmen." *Nat Neurosci* **18**, 800–806.

[10] Busche MA, Hyman BT (2020) Synergy between amyloid-β and tau in Alzheimer's disease. *Nat Neurosci* **23**, 1183–1193.

[11] Raj A, LoCastro E, Kuceyeski A, Tosun D, Relkin N, Weiner M (2015) Network Diffusion Model of Progression Predicts Longitudinal Patterns of Atrophy and Metabolism in Alzheimer's Disease. *Cell Rep* **10**, 359–369.

[12] Spires-Jones TL, Hyman BT (2014) The Intersection of Amyloid Beta and Tau at Synapses in Alzheimer's Disease. *Neuron* **82**, 756–771.

[13] Bertsch M, Franchi B, Marcello N, Tesi MC, Tosin A (2017) Alzheimer's disease: a mathematical model for onset and progression. *Math Med Biol J IMA* **34**, 193–214.

[14] Maccioni RB, Farías G, Morales I, Navarrete L (2010) The Revitalized Tau Hypothesis on Alzheimer's Disease. *Arch Med Res* **41**, 226–231.

[15] Hardy J (2009) The amyloid hypothesis for Alzheimer's disease: a critical reappraisal. *J Neurochem* **110**, 1129–1134.

[16] Selkoe DJ, Hardy J (2016) The amyloid hypothesis of Alzheimer's disease at 25 years. *EMBO Mol Med* **8**, 595–608.

[17] Maccioni RB, Muñoz JP, Barbeito L (2001) The Molecular Bases of Alzheimer's Disease and Other Neurodegenerative Disorders. *Arch Med Res* **32**, 367–381.

[18] Vogel JW, Iturria-Medina Y, Strandberg OT, Smith R, Levitis E, Evans AC, Hansson O (2020) Spread of pathological tau proteins through communicating neurons in human Alzheimer's disease. *Nat Commun* **11**, 2612.

[19] Hosseini-Asl E, Keynton R, El-Baz A (2016) Alzheimer's disease diagnostics by adaptation of 3D convolutional network. In *2016 IEEE International Conference on Image Processing (ICIP)*, pp. 126–130.

[20] Zhao X, Zhou F, Ou-Yang L, Wang T, Lei B (2019) Graph Convolutional Network Analysis for Mild Cognitive Impairment Prediction. In *2019 IEEE 16th International Symposium on Biomedical Imaging (ISBI 2019)*, pp. 1598–1601.

[21] Raj A, Kuceyeski A, Weiner M (2012) A Network Diffusion Model of Disease Progression in Dementia. *Neuron* **73**, 1204–1215.

[22] Iturria-Medina Y, Sotero RC, Toussaint PJ, Evans AC, Initiative and the ADN (2014) Epidemic Spreading Model to Characterize Misfolded Proteins Propagation in Aging and Associated Neurodegenerative Disorders. *PLOS Comput Biol* **10**, e1003956.

[23] Bertsch M, Franchi B, Tesi MC, Tosin A (2018) Well-posedness of a mathematical model for Alzheimer's disease. *ArXiv170905671 Math*.

[24] Hao W, Friedman A (2016) Mathematical model on Alzheimer's disease. *BMC Syst Biol* **10**, 108.

[25] Zhang Y, Hao Y, Li L, Xia K, Wu G, Alzheimer's Disease Neuroimaging Initiative (2020) A Novel Computational Proxy for Characterizing Cognitive Reserve in Alzheimer's Disease. *J Alzheimers Dis JAD* **78**, 1217–1228.

[26] Vicente E, Jha A, Frey E (2016) A nonparametric sinogram-based bootstrap resampling method to investigate scan time reduction in nuclear medicine imaging. *J Nucl Med* **57**, 1872–1872.

[27] Palop JJ, Chin J, Mucke L (2006) A network dysfunction perspective on neurodegenerative diseases. *Nature* **443**, 768–773.

[28] Destrieux C, Fischl B, Dale A, Halgren E (2010) Automatic parcellation of human cortical gyri and sulci using standard anatomical nomenclature. *NeuroImage* **53**, 1–15.

[29] Rowe CC, Ellis KA, Rimajova M, Bourgeat P, Pike KE, Jones G, Fripp J, Tochon-Danguy H, Morandeau L, O'Keefe G, Price R, Raniga P, Robins P, Acosta O, Lenzo

N, Szoeke C, Salvado O, Head R, Martins R, Masters CL, Ames D, Villemagne VL (2010) Amyloid imaging results from the Australian Imaging, Biomarkers and Lifestyle (AIBL) study of aging. *Neurobiol Aging* **31**, 1275–1283.

[30] Wolk DA, Price JC, Saxton JA, Snitz BE, James JA, Lopez OL, Aizenstein HJ, Cohen AD, Weissfeld LA, Mathis CA, Klunk WE, De-Kosky ST, DeKoskym ST (2009) Amyloid imaging in mild cognitive impairment subtypes. *Ann Neurol* **65**, 557–568.

[31] Hostetler ED, Walji AM, Zeng Z, Miller P, Bennacef I, Salinas C, Connolly B, Gantert L, Haley H, Holahan M, Purcell M, Riffel K, Lohith TG, Coleman P, Soriano A, Ogawa A, Xu S, Zhang X, Joshi E, Della Rocca J, Hesk D, Schenk DJ, Evelhoch JL (2016) Preclinical Characterization of 18F-MK-6240, a Promising PET Tracer for In Vivo Quantification of Human Neurofibrillary Tangles. *J Nucl Med Off Publ Soc Nucl Med* **57**, 1599–1606.

[32] Devanand DP, Mikhno A, Pelton GH, Cuasay K, Pradhaban G, Dileep Kumar JS, Upton N, Lai R, Gunn RN, Libri V, Liu X, van Heertum R, Mann JJ, Parsey RV (2010) Pittsburgh compound B (11C-PIB) and fluorodeoxyglucose (18 F-FDG) PET in patients with Alzheimer disease, mild cognitive impairment, and healthy controls. *J Geriatr Psychiatry Neurol* **23**, 185–198.

[33] Foster NL, Heidebrink JL, Clark CM, Jagust WJ, Arnold SE, Barbas NR, DeCarli CS, Turner RS, Koeppe RA, Higdon R, Minoshima S (2007) FDG-PET improves accuracy in distinguishing frontotemporal dementia and Alzheimer's disease. *Brain J Neurol* **130**, 2616–2635.

[34] Gong G, He Y, Chen ZJ, Evans AC (2012) Convergence and divergence of thickness correlations with diffusion connections across the human cerebral cortex. *NeuroImage* **59**, 1239–1248.

[35] Jack CR, Knopman DS, Jagust WJ, Shaw LM, Aisen PS, Weiner MW, Petersen RC, Trojanowski JQ (2010) Hypothetical model of dynamic biomarkers of the Alzheimer's pathological cascade. *Lancet Neurol* **9**, 119–128.

[36] Jack CR, Holtzman DM (2013) Biomarker modeling of Alzheimer's disease. *Neuron* **80**, 1347–1358.

[37] Frost GR, Li Y-M (2017) The role of astrocytes in amyloid production and Alzheimer's disease. *Open Biol* **7**, 170228.

[38] Goutelle S, Maurin M, Rougier F, Barbaut X, Bourguignon L, Ducher M, Maire P (2008) The Hill equation: a review of its capabilities in pharmacological modelling. *Fundam Clin Pharmacol* **22**, 633–648.

[39] Weiss JN (1997) The Hill equation revisited: uses and misuses. *FASEB J* **11**, 835–841.

[40] LaSalle J, Lefschetz S (1960) *Stability by Lyapunov's Second Method with Applications.*, Academic Press, New York.

[41] Hohman TJ, McLaren DG, Mormino EC, Gifford KA, Libon DJ, Jefferson AL, Initiative F the ADN (2016) Asymptomatic Alzheimer disease: Defining resilience. *Neurology* **87**, 2443–2450.

[42] Arenaza-Urquijo EM, Vemuri P (2018) Resistance vs resilience to Alzheimer disease: Clarifying terminology for preclinical studies. *Neurology* **90**, 695–703.

[43] Goldberg DE (1989) *Genetic Algorithms in Search, Optimization & Machine Learning*, Addison-Wesley.

[44] Pelikan M, Goldberg DE (1999) BOA: The Bayesian optimization algorithm. In *in Proc. Genetic and*, pp. 525–532.

[45] Kolda TG, Lewis RM, Torczon V (2003) Optimization by Direct Search: New Perspectives on Some Classical and Modern Methods. *SIAM Rev* **45**, 385–482.

[46] Evans DA, Hebert LE, Beckett LA, Scherr PA, Albert MS, Chown MJ, Pilgrim DM, Taylor JO (1997) Education and Other Measures of Socioeconomic Status and Risk of Incident Alzheimer Disease in a Defined Population of Older Persons. *Arch Neurol* **54**, 1399–1405.

[47] Karp A, Kåreholt I, Qiu C, Bellander T, Winblad B, Fratiglioni L (2004) Relation of Education and Occupation-based Socioeconomic Status to Incident Alzheimer's Disease. *Am J Epidemiol* **159**, 175–183.

[48] Fornari S, Schäfer A, Jucker M, Goriely A, Kuhl E (2019) Prion-like spreading of Alzheimer's disease within the brain's connectome. *J R Soc Interface* **16**, 20190356.

[49] Kaufman SK, Sanders DW, Thomas TL, Ruchinskas AJ, Vaquer-Alicea J, Sharma AM, Miller TM, Diamond MI (2016) Tau Prion Strains Dictate Patterns of Cell Pathology, Progression Rate, and Regional Vulnerability In Vivo. *Neuron* **92**, 796–812.

[50] Thal DR, Rüb U, Orantes M, Braak H (2002) Phases of Aβ-deposition in the human brain and its relevance for the development of AD. *Neurology* **58**, 1791–1800.

[51] Mattsson N, Palmqvist S, Stomrud E, Vogel J, Hansson O (2019) Staging β-Amyloid Pathology With Amyloid Positron Emission Tomography. *JAMA Neurol* **76**, 1319–1329.

[52] Schöll M, Lockhart SN, Schonhaut DR, O'Neil JP, Janabi M, Ossenkoppele R, Baker SL, Vogel JW, Faria J, Schwimmer HD, Rabinovici GD, Jagust WJ (2016) PET Imaging of Tau Deposition in the Aging Human Brain. *Neuron* **89**, 971–982.

[53] Cho H, Choi JY, Hwang MS, Lee JH, Kim YJ, Lee HM, Lyoo CH, Ryu YH, Lee MS (2016) Tau PET in Alzheimer disease and mild cognitive impairment. *Neurology* **87**, 375–383.

[54] Liu Q, Yang D, Zhang J, Wei Z, Wu G, Chen M (2021) Analyzing The Spatiotemporal Interaction And Propagation Of Atn Biomarkers In Alzheimer's Disease Using Longitudinal Neuroimaging Data. In *2021 IEEE 18th International Symposium on Biomedical Imaging (ISBI)*, pp. 126–129.

[55] Ossenkoppele R, Schonhaut DR, Schöll M, Lockhart SN, Ayakta N, Baker SL, O'Neil JP, Janabi M, Lazaris A, Cantwell A, Vogel J, Santos M, Miller ZA, Bettcher BM, Vossel KA, Kramer JH, Gorno-Tempini ML, Miller BL, Jagust WJ, Rabinovici GD (2016) Tau PET patterns mirror clinical and neuroanatomical variability in Alzheimer's disease. *Brain* **139**, 1551–1567.

[56] Veitch DP, Weiner MW, Aisen PS, Beckett LA, Cairns NJ, Green RC, Harvey D, Jack CR, Jagust W, Morris JC, Petersen RC, Saykin AJ, Shaw LM, Toga AW, Trojanowski JQ (2019) Understanding disease progression and improving Alzheimer's disease clinical trials: Recent highlights from the Alzheimer's Disease Neuroimaging Initiative. *Alzheimers Dement* **15**, 106–152.

[57] Mathis CA, Kuller LH, Klunk WE, Snitz BE, Price JC, Weissfeld LA, Rosario BL, Lopresti BJ, Saxton JA, Aizenstein HJ, McDade EM, Kamboh MI, DeKosky ST, Lopez OL (2013) In vivo assessment of amyloid-β deposition in nondemented very elderly subjects. *Ann Neurol* **73**, 751–761.

[58] Lerch JP, Pruessner JC, Zijdenbos A, Hampel H, Teipel SJ, Evans AC (2005) Focal decline of cortical thickness in Alzheimer's disease identified by computational neuroanatomy. *Cereb Cortex N Y N 1991* **15**, 995–1001.

[59] Franzmeier N, Rubinski A, Neitzel J, Kim Y, Damm A, Na DL, Kim HJ, Lyoo CH, Cho H, Finsterwalder S, Duering M, Seo SW, Ewers M (2019) Functional connectivity associated with tau levels in ageing, Alzheimer's, and small vessel disease. *Brain* **142**, 1093–1107.

[60] Guo Z, Zhang J, Liu X, Hou H, Cao Y, Wei F, Li J, Chen X, Shen Y, Chen W (2015) Neurometabolic characteristics in the anterior cingulate gyrus of Alzheimer's disease patients with depression: a 1H magnetic resonance spectroscopy study. *BMC Psychiatry* **15**, 306.

[61] Insel PS, Mormino EC, Aisen PS, Thompson WK, Donohue MC (2020) Neuroanatomical spread of amyloid β and tau in Alzheimer's disease: implications for primary prevention. *Brain Commun* **2**, fcaa007.

[62] Kumar S, Zomorrodi R, Ghazala Z, Goodman MS, Blumberger DM, Cheam A, Fischer C, Daskalakis ZJ, Mulsant BH, Pollock BG, Rajji TK (2017) Extent of Dorsolateral Prefrontal Cortex Plasticity and Its Association With Working Memory in Patients With Alzheimer Disease. *JAMA Psychiatry* **74**, 1266–1274.

[63] Cavedo E, Pievani M, Boccardi M, Galluzzi S, Bocchetta M, Bonetti M, Thompson PM, Frisoni GB (2014) Medial temporal atrophy in early and late-onset Alzheimer's disease. *Neurobiol Aging* **35**, 2004–2012.

[64] Gan CL, O'Sullivan MJ, Metzler-Baddeley C, Halpin S (2017) Association of imaging abnormalities of the subcallosal septal area with Alzheimer's disease and mild cognitive impairment. *Clin Radiol* **72**, 915–922.

[65] Buckner RL, Sepulcre J, Talukdar T, Krienen FM, Liu H, Hedden T, Andrews-Hanna JR, Sperling RA, Johnson KA (2009) Cortical Hubs Revealed by Intrinsic Functional Connectivity: Mapping, Assessment of Stability, and Relation to Alzheimer's Disease. *J Neurosci* **29**, 1860–1873.

[66] Guillon J, Attal Y, Colliot O, La Corte V, Dubois B, Schwartz D, Chavez M, De Vico Fallani F (2017) Loss of brain inter-frequency hubs in Alzheimer's disease. *Sci Rep* **7**, 10879.

[67] Wang Y, Yang D, Li Q, Kaufer D, Styner M, Wu G (2020) Characterizing the Propagation Pattern of Neurodegeneration in Alzheimer's Disease by Longitudinal Network Analysis. In *2020 IEEE 17th International Symposium on Biomedical Imaging (ISBI)*, pp. 292–295.

[68] Convit A, de Asis J, de Leon MJ, Tarshish CY, De Santi S, Rusinek H (2000) Atrophy of the medial occipitotemporal, inferior, and middle temporal gyri in non-demented elderly predict decline to Alzheimer's disease☆. *Neurobiol Aging* **21**, 19–26.

[69] Scheff SW, Price DA, Schmitt FA, Scheff MA, Mufson EJ (2011) Synaptic Loss in the Inferior Temporal Gyrus in Mild Cognitive Impairment and Alzheimer's Disease. *J Alzheimers Dis* **24**, 547–557.

[70] Cai H, Cong W, Ji S, Rothman S, Maudsley S, Martin B (2012) Metabolic Dysfunction in Alzheimer's Disease and Related Neurodegenerative Disorders. *Curr Alzheimer Res* **9**, 5–17.

[71] Zhang J, Yang D, He W, Wu G, Chen M (2020) A Network-Guided Reaction-Diffusion Model of AT[N] Biomarkers in Alzheimer's Disease. In *2020 IEEE 20th International Conference on Bioinformatics and Bioengineering (BIBE)*, pp. 222–229.

[72] Luo J, Wärmländer SKTS, Gräslund A, Abrahams JP (2016) Cross-interactions between the Alzheimer Disease Amyloid-β Peptide and Other Amyloid Proteins: A Further Aspect of the Amyloid Cascade Hypothesis. *J Biol Chem* **291**, 16485–16493.

[73] Xin S-H, Tan L, Cao X, Yu J-T, Tan L (2018) Clearance of Amyloid Beta and Tau in Alzheimer's Disease: from Mechanisms to Therapy. *Neurotox Res* **34**, 733–748.

[74] Zlokovic BV, Yamada S, Holtzman D, Ghiso J, Frangione B (2000) Clearance of amyloid β-peptide from brain: transport or metabolism? *Nat Med* **6**, 718–718.

# Supplementary Information

## Severely affected regions

Due to the space limitations, we provide the brain mapping of summarized severely affected regions here. **Amyloid:** (Fig. S1A) Mathis et al. conducted a longitudinal study, which investigated the effects of Ginkgo biloba (Gb) on preventing dementia [1]. They performed a control study to determine the elevation of amyloid-beta in multiple brain regions using MRI. The level of amyloid deposition is revealed by PiB retention, in which high PiB retention is correlated to high amyloid-beta concentration. Higher levels of PiB were found in anterior cingulate 6, 7, frontal cortex 20, 42, 19, 2, lateral temporal cortex 34, 37, 38, 43, parietal cortex 26, 27, 28, precuneus 30 in participants with mild cognitive impairment compared with normal control. The result from our model is consistent with these in vivo assessment findings.

**Tau:** (Fig. S1B) A wealth of preclinical *in vivo* and *in vitro* research in this decade implied that Tau proliferation in the brain follows a prion-like pattern transmitting between neurons. A study done by Franzmeier et al. assessed the relationship between functional connectivity and higher spatial Tau covariance in the normal control group and AD group [2]. Except for hippocampus, the result from their study indicated that Tau was also attacking transentorhinal cortex, entorhinal cortex 32, and medial temporal limbic system heavily (all other regions). Regional Tau-PET levels revealed a network-specific profile of tau distribution, with Tau highly aggregating in medial temporal limbic network. Other cortical networks such as sensory areas and association areas also exhibited an increase in Tau, though motor network was devoid for the most part. In Figure B, 32 corresponds to the transentorhinal and entorhinal cortex, and 42, 11, 44 corresponds to the medial temporal limbic system, both are regions with early tau attack, and these are well confirmed by these references.

**Neurodegeneration:** (Fig. S1C) The key nodes of neurodegeneration are determined by cortical thickness. Regions with a drastic decline in cortical thickness demonstrate a high neuropathological alternation caused by AD. Lerch et al. identified left anterior cingulate 6, 7, dorsolateral prefrontal cortex 14, orbitofrontal cortex 24, visual association cortex 20, 25, and medial temporal lobe 32, 23, outlining parahippocampal gyrus 23 as the most prominent regions under AD's attack [3].

We also compared the averaged AT[N] level at each in the corresponding heatmaps (Supplementary Fig. S1D-F), the overall increase of neuropathological burdens at vulnerable regions are higher than regions summarized in literature, especially for tau, confirming the susceptibility of vulnerable nodes and significant increases of tau level in AD development. Note that the heatmap only represents the SUVR levels at vulnerable and summarized brain regions.

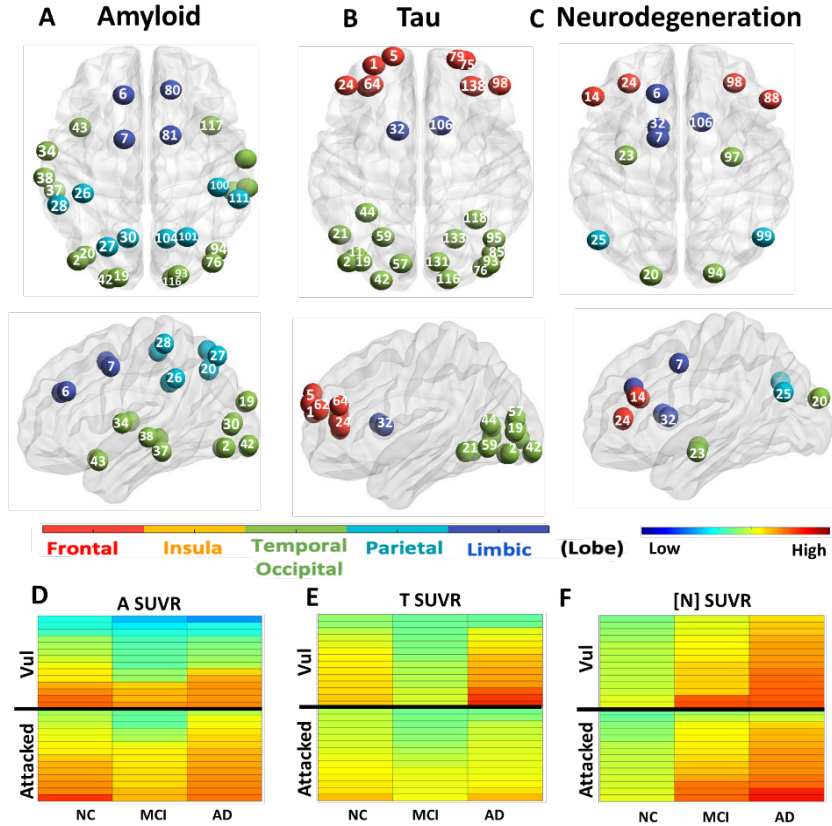

**Figure S1. Brain mapping of severely affected regions for amyloid, tau, and neurodegeneration from literature.** (**A**)-(**C**) Brain mapping of Vulnerable regions detected by our model. Ball colors indicate different brain lobes. Node ID is labeled in white, referring to region name list in Figure 4A. Reversed FDG is used as an indicator for [N] biomarker. (**D**)-(**F**) Heatmaps of AT[N] longitudinal changes between last scan and first scan for NC, MCI, and AD groups. Top 14 severely attacked regions and vulnerable brain regions are shown in comparison and sorted by the difference of SURV level in ascending order.

# REFERENCES

[1]  Mathis CA, Kuller LH, Klunk WE, Snitz BE, Price JC, Weissfeld LA, Rosario BL, Lopresti BJ, Saxton JA, Aizenstein HJ, McDade EM, Kamboh MI, DeKosky ST, Lopez OL (2013) In vivo assessment of amyloid-β deposition in nondemented very elderly subjects. *Ann Neurol* **73**, 751–761.
[2]  Franzmeier N, Rubinski A, Neitzel J, Kim Y, Damm A, Na DL, Kim HJ, Lyoo CH, Cho H, Finsterwalder S, Duering M, Seo SW, Ewers M (2019) Functional connectivity associated with tau levels in ageing, Alzheimer's, and small vessel disease. *Brain* **142**, 1093–1107.
[3]  Lerch JP, Pruessner JC, Zijdenbos A, Hampel H, Teipel SJ, Evans AC (2005) Focal decline of cortical thickness in Alzheimer's disease identified by computational neuroanatomy. *Cereb Cortex N Y N 1991* **15**, 995–1001.